# Supervised Ranking of Triples for Type-Like Relations

The Cress Triple Scorer at the WSDM Cup 2017

Faegheh Hasibi[1]    Darío Garigliotti[2]    Shuo Zhang[2]    Krisztian Balog[2]

[1]NTNU Trondheim
faegheh.hasibi@idi.ntnu.no

[2]University of Stavanger
<firstname.lastname>@uis.no

## ABSTRACT
This paper describes our participation in the Triple Scoring task of WSDM Cup 2017, which aims at ranking triples from a knowledge base for two type-like relations: profession and nationality. We introduce a supervised ranking method along with the features we designed for this task. Our system has been top ranked with respect to average score difference and 2nd best in terms of Kendall's tau.

## 1. INTRODUCTION

The *Triple Scoring* task of WSDM Cup 2017 aims at computing relevance scores for triples from a knowledge base for type-like relations [5, 2]. For example, considering *profession* as a type-like relation, the objective of this task is to generate the following triple scores:

| | | | |
|---|---|---|---|
| *William Shakespeare* | *has-profession* | *Poet* | 7 |
| *William Shakespeare* | *has-profession* | *Lyricist* | 3 |
| *Bob Dylan* | *has-profession* | *Singer-songwriter* | 7 |
| *Bob Dylan* | *has-profession* | *Guitarist* | 5 |

More specifically, the task is defined as: "Given a list of triples from two type-like relations (profession and nationality), for each triple compute an integer score from 0..7 that measures the degree to which the subject belongs to the respective type (expressed by the predicate and object)" [5]. Such scores can be used to identify the main professions of a person (e.g., "Bob Dylan"), or can be used in the ranking of entity-bearing queries like "american poets" [1].

We address this task using a supervised ranking method, where a separate ranking function is learned for each type-like relation. The ranking function generates scores for subject-object pairs, each represented as a set of features extracted from Wikipedia. The task is evaluated using three evaluation metrics: average score difference, Kendall's tau coefficient, and accuracy. According to the official WSDM Cup results, our approach is the top performing system with respect to average score difference, 2nd best with respect to Kendall's tau, and 6th overall in terms of accuracy.

In the remainder of this paper, we detail our approach in Section 2 and present our experimental results in Section 3.

## 2. APPROACH

We approach the *Triple Scoring* task using a supervised ranking method. We aim to learn a ranking function, which can generate scores for person-profession or person-nationality pairs. Each pair is represented by a set of features, extracted from Wikipedia text snippets. Below, we describe the data sources used for extracting features (Section 2.1), the description of the features for each relation (Sections 2.2 and 2.3), and the machine learning method employed (Section 2.4). We refer to Table 2 for the notation used in this paper.

Table 2: Glossary of notations.

| Symbol | Description |
|---|---|
| $t$ | Term (word) |
| $pe$ | Person |
| $pr$ | Profession |
| $nt$ | Nationality |
| $s$ | Wikipedia sentence |
| $|S|$ | Total number of Wikipedia sentences |
| $t \in s$ | The sentence $s$ contains the term $t$ |
| $Pr(pe)$ | Set of professions of person $pe$ |
| $T_k(pr)$ | Set of top-$k$ terms for profession $pr$ |
| $S(p)$ | Wikipedia sentences of $p \in \{pr, pe\}$ |
| $tf(t,s)$ | Raw term frequency of $t$ in $s$ |
| $w(t,pe)$ | Term weight for a person |
| $w(t,pr)$ | Term weight for a profession |
| $vec(t)$ | 300-dim. word embedding vector of $t$ |

### 2.1 Data sources

We rely on Wikipedia as the main source for extracting features. A set of 33,159,353 Wikipedia sentences, annotated with 385,426 entities, was provided as part of the task. We built an index from these sentences, where annotations were replaced by the respective entity IDs (provided by the task organizers). For example, the text snippet ".. directed by [Ventura_Pons|Ventura Pons]" is indexed as ".. directed by fb_06dfpq".[1] The index contains an additional field for storing the professions of all entities mentioned in the sentence. We utilize this index to extract statistics for the profession relation.

Another source of data used for both profession and nationality relations is the first sentences from each person's Wikipedia article. These sentences, particularly the first one, often mention the most important professions/nationality of a person. For convenience, we extracted the first sentence and paragraph from DBpedia (version 2015-10), via the dbo:abstract and rdfs:comment predicates.

### 2.2 Profession features

Our general strategy is to construct a term-based representation of persons and professions from the Wikipedia sentences that are made available as part of the challenge. We write $w(t, \cdot)$ to denote the weight of term $t$ using some term weighting scheme, which will always be TF.IDF in this work. Let $S(pe)$ be the set of sentences that mention person $pe$: $S(pe) = \{s : pe \in s\}$. We further know from the knowledge base (KB) the set of professions that a person has, $Pr(pe)$. Putting the two together, we obtain a set of sentences for each profession, i.e., sentences that mention a person that has that profession: $S(pr) = \{s : pe \in s \land pr \in Pr(pe)\}$.

---
[1]The Freebase ID m.06dfpq is converted to fb_06dfpq to avoid undesired segmentation of entity IDs by the text analyzer.

**Table 1:** Features used for scoring triples from profession and nationality relations. Features involving a parameter $k$ are generated for $k \in [10, 50, 100, 200, 500, 1000]$.

| Profession features | | Value |
|---|---|---|
| sumProfTerms{k} | The sum of top-$k$ (TF.IDF weighted) profession terms from all Wikipedia sentences of the person | $[0, \inf)$ |
| simCos{k} | Cosine similarity between the profession and person (TF.IDF weighted) term vectors of top-$k$ profession terms | $[0, 1]$ |
| simCosVec{k} | Cosine similarity between the profession and person embedding vectors (centroids of TF.IDF weighted Word2Vec vectors) | $[0, 1]$ |
| simCosVecPar{100} | Cosine similarity between the profession and person embedding vectors based on the first Wikipedia paragraph of a person | $[0, 1]$ |
| isProfWPSent | Whether the profession occurs in the first Wikipedia sentence of the person | $\{0, 1\}$ |
| isProfWPPar | Whether the profession occurs in the first Wikipedia paragraph of the person | $\{0, 1\}$ |
| isFirstProfWPSent | Whether the profession is the first of the professions occurring in the first Wikipedia sentence of the person | $\{0, 1\}$ |
| isFirstProfWPPar | Whether the profession is the first of the professions occurring in the first Wikipedia paragraph of the person | $\{0, 1\}$ |
| **Nationality features** | | |
| freqPerNat | The normalized frequency of a person co-occurring with a nationality in Wikipedia sentences | $[0, inf)$ |
| isNatWPSent | Whether the nationality occurs in the first Wikipedia sentence of the person | $\{0, 1\}$ |
| isNatWPPar | Whether the nationality occurs in the first Wikipedia paragraph of the person | $\{0, 1\}$ |
| isFirstNatWPSent | Whether the nationality is the first of the nationalities occurring in the first Wikipedia sentence of the person | $\{0, 1\}$ |
| isFirstNatWPPar | Whether the nationality is the first of the nationalities occurring in the first Wikipedia paragraph of the person | $\{0, 1\}$ |

To build the term-based representation of a profession, we take all the terms associated with a profession and rank them based on a weighting scheme; that is for each term in the sentences of $S(pr)$, we compute the term weight $w(t, pr)$ as:

$$w(t, pr) = \text{TFIDF}(t, pr) = \frac{\sum_{s \in S(pr)} tf(t,s)}{\sum_{s \in S(pr)} |s|} \log \frac{|S|}{|\{s : t \in s\}|}.$$

This weight reflects the importance of a term for a given profession. In similar vein, the term weight for a person is computed as:

$$w(t, pe) = \text{TFIDF}(t, pe) = \frac{\sum_{s \in S(pe)} tf(t,s)}{\sum_{s \in S(pe)} |s|} \log \frac{|S|}{|\{s : t \in s\}|}.$$

To obtain high quality term representations, we perform further cleansing by filtering out terms that are entity IDs or stopwords (using a set of 127 stopwords).

Given these basic notations, we now turn into describing the individual features presented in top part of Table 1.

**Sum of profession terms (sumProfTerms{k}):** Takes the top-$k$ terms of a profession and computes the weighted sum of term frequencies from all Wikipedia sentences of a person:

$$\text{sumProfTerms}_k(pe, pr) = \sum_{t \in T_k(pr)} \left( w(t, pr) \sum_{s \in S(pe)} tf(t, s) \right).$$

**Cosine similarity (simCos{k}):** Computes the cosine similarity between the person's and profession's term vectors:

$$\text{simCos}_k(\vec{t}_{pe}^{\,k}, \vec{t}_{pr}^{\,k}) = \frac{\sum_{t \in T_k(pr)} w(t, pe) w(t, pr)}{||\vec{t}_{pe}^{\,k}||\,||\vec{t}_{pr}^{\,k}||},$$

where $\vec{t}_{pr}^{\,k}$ and $\vec{t}_{pe}^{\,k}$ are $k$-dimensional vectors, holding weights $w(t, pr), w(t, pe)$ for the each term $t \in T_k(pr)$.

**Word2vec cosine similarity (simCosVec{k}):** Measures the cosine similarity between a person and a profession based on the embedding vectors; i.e., 300 dimensional Word2Vec vectors trained on the Google news dataset [6]:

$$\text{simCosVec}_k(\vec{C}_{pe}^{\,k}, \vec{C}_{pr}^{\,k}) = \frac{\vec{C}_{pe}^{\,k} \cdot \vec{C}_{pr}^{\,k}}{||\vec{C}_{pe}^{\,k}||\,||\vec{C}_{pr}^{\,k}||}, \quad (1)$$

where $\vec{C}_{pr}^{\,k}, \vec{C}_{pr}^{\,k}$ represent the centroid vectors of a person and profession, respectively, and are computed based on the TF.IDF weighted sum of the top-$k$ term vectors:

$$\vec{C}_{pr}^{\,k} = \sum_{t \in T_k(pr)} w(t, pr) vec(t), \quad (2)$$

$$\vec{C}_{pe}^{\,k} = \sum_{t \in T_k(pr)} w(t, pe) vec(t). \quad (3)$$

**Word2vec cosine similarity (simCosVecPar{100}):** This feature, similar to the previous one, measures the similarity between a person and profession using the Word2Vec vectors. Here, however, the vector representation of the person is built from the terms in the first Wikipedia paragraph of the person $pe$:

$$\vec{C}_{pe}^{\,par} = \sum_{t \in par_{pe}} tf(t, par_{pe}) vec(t),$$

Table 3: Triple scoring results for profession and nationality relations on both train and test collection.

| Model | Accuracy | ASD | Kendall |
|---|---|---|---|
| Profession (train) | 0.77 | 1.61 | 0.34 |
| Profession (test) | 0.78 | 1.61 | 0.27 |
| Nationality (train) | 0.76 | 1.62 | 0.32 |
| Nationality (test) | 0.77 | 1.62 | 0.40 |

where the each term $t$ is of at least length 4 and appeared more than one time in the Wikipedia paragraph (i.e., $|t| \geq 4$ and $tf(t, par_{pe}) \geq 2$). The cosine similarity is then computed according to Eq. (1) and between vectors $\vec{C}_{pe}^{par}$ and $\vec{C}_{pr}^{100}$, the latter being computed using Eq. (2).

**Other features:** The four other profession features presented in Table 1 are meant to capture the most important profession of a person. The features **isProfWPSent** and **isProfWPPar** check whether the profession appears in the first Wikipedia sentence/paragraph of a person, and **isProfWPSent** and **isProfWPPar** examine whether the profession is the first among the mentioned professions in the corresponding text.

### 2.3 Nationality features

We consider two forms of nationality and compute features for each of these forms: (1) when nationality is used as an adjective (e.g., German), and (2) when it is referred to as a country name, i.e., is a noun (e.g., Germany). The bottom part of Table 1 summarizes the nationality features used in our learning method.

**Person-nationality frequency (freqPerNat):** The feature is specified as the number of times a nationality appears in the Wikipedia sentences of a person, normalized by the total number of sentences in which the person occurs:

$$\text{freqPerNat}(pe, nt) = \frac{|\{s : pe \in s, nt \in s\}|}{|S(pe)|}. \qquad (4)$$

**Other features:** The other features we employ for the nationality relation are extracted from the first paragraph/sentence of a person's Wikipedia article. These are similar to the same group of profession features, except that we search for nationality terms (both adjective and noun forms) in the text.

### 2.4 Generating triple scores

For each relation, a set of labeled subject-object pairs was provided as the input (training) data. Each pair (i.e., person-profession or person-nationality) represents a learning instance that has to be scored. We employ the Random Forest (RF) regressor [3] as our supervised ranking method and generate two separate models for the profession and nationality relations. We trained these models on the labeled data provided by the task organizers and generate a score for each subject-object pair, represented by a feature vector. The Random Forest algorithm involves two parameters: number of trees (iterations) $n$, and maximum number of features in each tree $m$. We set $n = 1000$ for both models, $m = 3$ for profession, and $m = 2$ for nationality model (around 10% of the feature set [4]).

Once the triples are ranked using the ranking model, we need to map them to integer scores in the range 0..7 [2]. To this end, we round the ranking score to the closest integer number. More

Table 4: Top performing systems according to the official WSDM Cup evaluation results. The best scores for each metric are presented in boldface. *Rank* refers to the relative ordering of systems according to the given metric.

| | cress [our] | bokchoy | goosefoot |
|---|---|---|---|
| Accuracy | 0.78 | **0.87** | 0.75 |
| rank | 6 | 1 | >7 |
| ASD | **1.61** | 1.63 | 1.78 |
| rank | 1 | 2 | 7 |
| Kendall | 0.32 | 0.33 | **0.31** |
| rank | 2 | 3 | 1 |

specifically, we employ the following function to convert the ranking score $s_r$ to a triple score:

$$f(s_r) = \begin{cases} round(s_r), & 0 \leq s_r \leq 7 \\ 0, & s_r < 0 \\ 7, & s_r > 7. \end{cases} \qquad (5)$$

## 3. EXPERIMENTAL RESULTS

In this section, we describe the evaluation settings, present the results of our system, and analyze the features used for the profession and nationality relations.

### 3.1 Evaluation

We evaluate our system on two collections: *training* and *test*. The training set contains 515 and 162 instances for profession and nationality relations, respectively, and was provided as part of the task [2, 5]. We report on the results of this test collection using 5-fold cross validation. The test collection was used by the task organizers to evaluate all the submissions and was made publicly available only after the submission deadline; it consists of 513 and 197 instances for profession and nationality relations, respectively. We use the entire training set for learning when generating results for the test collection.

System performance is evaluated using three relevance metrics: (i) accuracy, (ii) average score difference (ASD), and (iii) Kendall's tau. The first two metrics are based on the absolute scores values, while the last one considers the relative ranking of triples. High accuracy, and low ASD and Kendall are desired.

### 3.2 Results

**Triple scoring results.** Table 3 presents the results of our system for profession and nationality, on the training and test collections. The results show that our models deliver good results and are comparable to the best approaches of Bast et al. [1]. The results also reveal that the rank-based metrics (accuracy and ASD) are highly correlated across different relations and collections, which is not the case for Kendall's tau.

**WSDM Cup evaluation results.** Table 4 presents the results of the top performing systems with respect to each of the evaluation metrics (according to the official task results[2]). These numbers indicate the overall performance of systems, i.e., both relations (profession and nationality) combined. Our system, denoted as "cress," is ranked first and second with respect to ASD and Kendall, respectively, and is sixth in terms of to accuracy. Note that this is unlike the results in Table 3, where accuracy and ASD are correlated. As explained by the task organizers, some groups truncated their scores to the range 2..5, which made accuracy better, but ASD and Kendall worse. When factoring out the effect of the 2..5 truncation trick, the rankings according to all metrics are highly correlated. Further details on result comparison are provided in [2].

---

[2] http://www.tira.io/task/triple-scoring/dataset/wsdmcup17-triple-scoring-test-dataset2-2016-12-08/

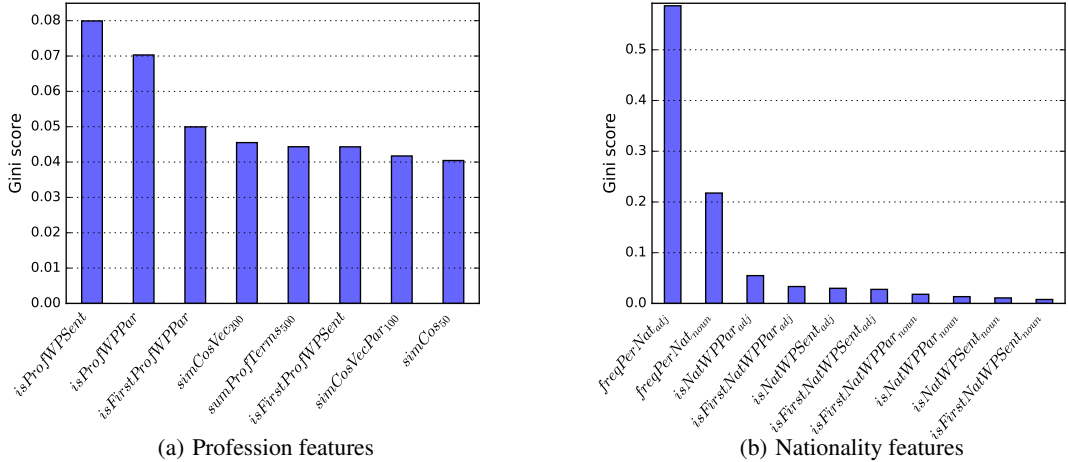

Figure 1: Feature importance for the profession and nationality relations.

### 3.3 Feature analysis

We now analyze the features we developed for the two relations.

**Profession features.** Figure 1(a) presents the importance of features for the profession relation according to their Gini score. For the features involving a $k$ parameter (the first three features of Table 1), we only kept the highest performing variant. We observe that that features extracted from the first Wikipedia sentence/paragraph are of the highest importance. The other features, extracted from the Wikipedia sentences of persons, have approximately the same importance.

**Nationality features.** Figure 1(b) shows the feature importance for the nationality relation. It is immediately noticeable that the "freqPerNat" feature plays the most important role (in both adjective and noun variants). The third most important feature, "isNatWPPar," checks the presence of nationality in the first Wikipedia paragraph a person. The figure also reveals that the adjective forms of nationalities have greater impact than the noun forms.

In summary, our feature importance analysis highlights that the triple scoring task can benefit not only from a large collection of person-annotated Wikipedia sentences, but also from the first sentences of a person's Wikipedia article, even in the form of very basic features.

### 4. CONCLUSIONS

In this paper we have presented our participation in the Triple Scoring task of WSDM 2017 Cup, where our system has been one of the top performers. We have employed a supervised learning approach with a set of novel features designed specifically for the task at hand. We have reported on the results of our system, on both the training and test collections, and have complemented it with a feature analysis.